\documentclass{article}
\usepackage{RR}
\RRNo{6383}
\usepackage{hyperref}
\usepackage{graphicx,amssymb,amsmath,color}
\usepackage{subfigure}
\usepackage{times}
\usepackage{psfrag}
\usepackage{rotating}
\usepackage{latexsym}
\usepackage{amsbsy}
\usepackage{amsgen}
\usepackage{amsfonts}
\usepackage[english]{babel}

\newcommand{\vectornorm}[1]{\left|\left|#1\right|\right|}
\newcommand{\argmin}{\mathop{\mathrm{arg\,min}}}
\newcommand{\ddiv}{\;\mathbf{div} \,}
\newcommand{\dgrad}{\;\mathbf{grad} \,}

\newcommand{\vect}[1]{\mathbf{#1}}

\newcommand{\ct}{\textup{cost}}
\RRdate{D\'ecembre 2007}
\RRauthor{
  Marcelo Buffoni\thanks[sfr]{INRIA Futurs, Equipe-Projet MC2 and Institut de
  Math\'ematiques de Bordeaux, UMR 5251 CNRS,
  Universit\'{e} Bordeaux 1, 33405 Talence cedex, France.} 
  \and
  Haysam Telib\thanks{Dipartimento di Ingegneria Aeronautica e Spaziale,
  Politecnico di Torino. 10129 Torino, Italy}
  \and
  Angelo Iollo\thanksref{sfr}
}
\authorhead{M. Buffoni \& H. Telib \& A. Iollo}
\RRtitle{M\'ethodes It\'eratives pour la R\'eduction de Mod\`eles
  par D\'ecomposition de Domaine}  
\RRetitle{Iterative Methods for Model Reduction by Domain Decomposition}
\titlehead{Iterative Methods for Model Reduction by Domain Decomposition}
\RRresume{On propose une m\'ethode pour r\'eduire les efforts de calcul pour
r\'esoudre une \'equation aux d\'eriv\'ees partielles sur un
domaine donn\'e. L'id\'ee principale est de diviser le domaine consid\'er\'e en
deux sous-domaines, et d'employer diff\'erentes m\'ethodes d'approximation
dans chacun des deux sous-domaines. En particulier, dans un des sous-domaines
l'\'equation en question est discr\'etis\'ee par une m\'ethode canonique, tandis
que dans l'autre un mod\`ele d'ordre r\'eduit du probl\`eme original est
utilis\'e. Des strat\'egies diff\'erentes pour coupler le mod\`ele
d'ordre r\'eduit \'a la discr\'etisation habituelle sont
pr\'esent\'es. L'efficacit\'e de ces approches est test\'ee sur
des exemples num\'eriques pertinentes pour des probl\`emes mod\`eles non
lin\'eaires, notamment l'\'equation de Laplace, avec des conditions
limites non lin\'eaires, et les \'equations d'Euler compressibles.}

\RRabstract{We propose a method to reduce the computational effort to
  solve a partial differential equation on a given domain. The main
  idea is to split the domain of interest in two subdomains, and to
  use different approximation methods in each of the two
  subdomains. In particular, in one subdomain we discretize the
  governing equations by a canonical scheme, whereas in the other one
  we solve a reduced order model of the original problem. Different
  approaches to couple the low-order model to the usual
  discretization are presented. The effectiveness of these approaches
  is tested on numerical examples pertinent to non-linear model
  problems including the Laplace equation with non-linear
  boundary conditions and the compressible Euler equations.}  
\RRmotcle{mod\`eles r\'eduits,  d\'ecomposition de domaine, \'ecoulements compressibles}  
\RRkeyword{low-order models, domain decomposition, compressible flows}  
\RRprojet{MC2}
\RRtheme{\THNum} 
\URFuturs 
\begin{document}
\makeRR   

\section{Introduction}
\label{sec:1}
In this contribution we are concerned with the coupling between a full
order simulation and a reduced order model. The idea is to reduce the
extent of the domain where we perform a canonical numerical simulation
by introducing a low-order model which describes the solution far from
the region of interest. By reducing the extent of the domain we aim at
reducing the costs in terms of required memory as well as in terms of
computational time.   

In a broad sense, there exist many applications where far from the
boundary the solution is weakly dependent on the details of the
boundary geometry. In such regions we use a reduced order model based
on proper orthogonal decomposition (POD) \cite{L67} to solve the
problem. This approach allows a representation of the solution by a
small number of unknowns that are the coefficients of an appropriate
Galerkin expansion. Therefore away from a narrow region close to the
boundary of interest the number of unknowns to be solved for is
drastically reduced. This idea was previously explored  
in the context of transonic flows 
with shocks \cite{lucia}, \cite{alonso}. Here we extend those works by
adapting to that context some classical domain decomposition techniques.

In
the following we discuss three possible methods to do that. The first
is based on a Schur iteration where the solution of the low-order
model is obtained by a projection step in the space spanned by the POD
modes. The second is  in the same spirit but instead of a
Dirichlet-Neumann iteration we employ a Dirchlet-Dirchlet iteration in
the frame of a classical Schwarz method. The last approach is of
different nature since the solution of the low-order model is not
simply based on a projection in the space of the POD modes. It 
takes into account in a weak sense the governing equations by
minimizing the residual norm of the canonical approximation 
in the space spanned by the POD modes.

The numerical demonstrations shown in the following are relative to
two models: the Laplace equation with non-linear boundary conditions
modeling radiative heat transfer and the compressible Euler equations
in a nozzle. Since this method can be of interest for optimal design
applications, where many different geometries must be tested to
improve performance, in some cases we have explored the idea of
simulating by usual discretization methods just the region where the
geometry changes, modeling the rest by POD.       

Like all other approaches based on POD, a
solution database is necessary to build the basis functions, therefore
this method will be useful when many computations for relatively
similar cases are to be performed, like for example in shape
optimization, see  \cite{willcox}.

\section{Solution by projection of the solution trace in the space spanned by
  the POD modes}  
\label{sec:2}

\subsection{Approximation of the Steklov-Poincar\'e operator by POD}

\begin{figure}[h!]
  \centering 
  \includegraphics[width=7.cm,height=7.cm]{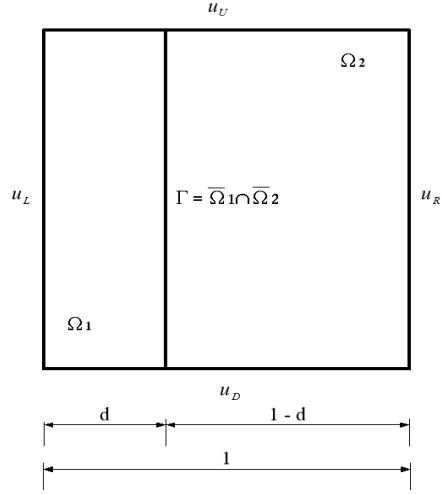} 
  \caption{{\em Problem set-up: Schur}}
  \label{fig_dom_a}
\end{figure}

In order to explain the method we take a particular case. Let us
consider the Laplace equation $\Delta u=0$ defined inside a square
$\Omega=[0,1]\times[0,1]$. Let $d\in[0,1]$,
$\Omega_1=[0,d]\times[0,1]$, $\Omega_2=[d,1]\times[0,1]$ and
$\Gamma=\overline{\Omega}_1 \cap \overline{\Omega}_2$ the interface
between the two sub-domains, see fig.~\ref{fig_dom_a}. We have
Dirichlet conditions on the right boundary ($u_R$) as well as on the
upper ($u_U$) and lower ($u_D$)
boundaries.  

We want to solve this problem for different values of the Dirichlet
data on the left ($u_L$) boundary. To that end we build an appropriate
solution database for a given set of   boundary conditions on  the left
side. In particular, the Dirichlet data on the left boundary is
denoted by $g^k$,
$1 \leq k \leq N$.  

Let the functions $u_\Omega^{(k)}$ be the discrete solutions of the
boundary value problem posed in $\Omega$ for different $k$, and let
$\bar{u}_2$ be an harmonic function, restricted to $\Omega_2$, defined
as follows: $\bar{u}_2=1/N \sum_{k=1}^N u_\Omega^{(k)}$. We compute a
Galerkin base of the form $\phi_i=\sum_{k=1}^{N} b_{i k}
(u_2^{(k)}-\bar{u}_2)$ where $u_2^{(k)}$ is the restriction of
$u_\Omega^{(k)}$ to $\Omega_2$. The coefficients $b_{ik}$ are found
by POD as explained in \cite{S87}. 
It can be shown that this 
base gives by construction an optimal representation of the
original data set $u_\Omega^{(k)}$.  

Let us define $\hat{u}_2=\bar{u}_2+\sum_{i=1}^{M} a_{i} \phi_i$, where
$M$ is much smaller than the number of discretization points in
$\Omega_2$. For an arbitrary Dirichlet condition on the left boundary
of $\Omega$, we want to determine the discrete solution by a canonical 
approximation in $\Omega_1$ and by the above defined Galerkin
representation in $\Omega_2$.   

One simple way to implement this idea is to solve the problem by 
Dirichlet-Neumann iterations. To this end, we follow the steps below:  
{\sf
\begin{enumerate}
\item
  solve the problem in $\Omega_1$ by any discretization method (FD, FEM,
  etc.), imposing Neumann b.c. on $\Gamma$; 
\item
  on interface $\Gamma$ project the trace of the above solution in the
  subspace spanned by the traces of the POD modes $\phi_i$; 
\item
  recover $\hat{u}_2$ as the prolongation of the trace of $\hat{u}_2$
  on $\Gamma$ inside $\Omega_2$ by using the POD modes;
\item
  set $\partial \hat{u}_1/\partial n=\partial \hat{u}_2/\partial n$ on $\Gamma$;
\item
  goto (1) until convergence is attained.
\end{enumerate}
}
This is just one possible solution algorithm, corresponding to a
classical domain decomposition method (Schur complement). Another
approach consists in solving the problem all at once, as detailed in
the following. Let us define $A_1$ the discretized operator acting on
$u_1$, the restriction of the unknowns belonging to $\Omega_1$;
$A_\Gamma$ the discretized operator acting on $u_\Gamma$, the unknowns
belonging to $\Gamma$ and $A_2$ the discretized operator acting on
$u_2$, the restriction of the unknowns belonging to $\Omega_2$.  

The discretized non-linear problem in $\Omega$ can be written 
\begin{equation}
\left(    
\begin{array}{lll}
A_1   &   B_1      &    0     \\
B_1^t &   A_\Gamma &    B_2^t \\
0     &   B_2      &    A_2   
\end{array}
\right) 
\left(
\begin{array}{l}
u_1      \\
u_\Gamma \\
u_2
\end{array}
\right)
=
\left(
\begin{array}{l}
f_1      \\
f_\Gamma \\
f_2
\end{array}
\right)
\label{eqschur}
\end{equation}
where $B_1$ and $B_2$ are appropriate interface matrices and $f_1$,
$f_\Gamma$, $f_2$ take into account the boundary conditions.  

From (\ref{eqschur}) we have 
\begin{eqnarray} \nonumber  
A_1 u_1+B_1 u_\Gamma=f_1 \\ 
\label{eqschur1}
B_1^t u_1+(A_\Gamma-B_2^t A_2^{-1} B_2) u_\Gamma=f_\Gamma-B_2^t
A_2^{-1}f_2 
\end{eqnarray}
The matrix $A_\Gamma-B_2^t A_2^{-1} B_2$ is the discrete counterpart
of the Steklov-Poincar\'e operator for $\Omega_2$, see \cite{Q99}.  
Consider now the second step of the solution algorithm proposed
above. Let $a\in\mathbb{R}^M$ be a vector of components $a_1 \dots
a_M$ and $c\in\mathbb{R}^M$ a vector of components $c_1 \dots
c_M$. Posing $\varphi_i$ the trace of $\phi_i$ on $\Gamma$, we take  
\begin{equation}
a = \argmin_{c \in \mathbb{R}^M}
\left(\;\;\vectornorm{u_\Gamma-\sum_{k=1}^M c_k     \varphi_k}\;\right) 
\label{eqproject}
\end{equation}
where $\vectornorm{\cdot}$ is the norm induced by the canonical $l^2$
scalar product, noted by $(\cdot,\cdot)$. 

Solution of (\ref{eqproject}) reduces to the solution  of the linear
problem $\sum_{i=1}^M a_i
\left(\varphi_i,\varphi_j\right)=(u_\Gamma,\varphi_j)$, $1\leq j \leq
M$.    
Therefore $a_i=(u_\Gamma,P_i)$,  where $P_i= \sum_{j=1}^M
\left[\left(\varphi_i,\varphi_j\right)\right]^{-1} \varphi_j$ is a
constant vector computed once for all from the POD modes. 

At this point we approximate $u_2$ with $\hat{u}_2$ and substitute in
(\ref{eqschur}). Since $B_2^t \hat{u}_2=B_2^t \bar{u}_2+\sum_{i=1}^{M}
a_{i} B_2^t \phi_i$, we have $B_2^t \hat{u}_2 =B_2^t \bar{u}_2 +
\sum_{i=1}^M B_2^t \phi_i (u_\Gamma,P_i)$. Finally, letting
$\hat{S}_2=\sum_{i=1}^M B_2^t \phi_i P_i $ we obtain the approximation
of (\ref{eqschur1})  
\begin{equation}
B_1^t u_1+(A_\Gamma-\hat{S}_2) u_\Gamma=f_\Gamma-B_2^t \bar{u}_2 
\label{appstek}
\end{equation} 
where $B_2^t \bar{u}_2 \equiv B_2^t A_2^{-1}f_2$. Matrix $\hat{S}_2$
is the approximation of the discrete Steklov-Poincar\'e operator
obtained by the POD expansion. Equations (\ref{eqschur1}) can of
course be solved simultaneously by a standard linear solver. 

Just like for the usual Steklov-Poincar\'e operator, (\ref{appstek}) 
amounts to a non-local boundary condition for the problem posed in 
$\Omega_1$. The main advantage of this approach compared to computing
explicitly $S_2$ is that we do not need $A_2^{-1}$ to build
$\hat{S}_2$.  

In the following we present some numerical applications. The first
case is a plain application of the Schur complement as it was
explained above to a non-linear case. The second case is based on
the Schwarz method. The  third numerical experience is a 
variant of the all at one method.

\subsubsection{Schur complement}
\label{schu}

A second order finite differences (FD) method coupled to a fix point
iteration is used to solve the Laplace equation inside the square
domain shown in fig.~\ref{fig_dom_a}. The left Dirichlet boundary
condition is varied to build the needed database. In particular,
$u_L=\sin{(k \pi y)+y}$, $1 \leq k \leq 49$. The boundary conditions
on the other sides are, referred to fig.~\ref{fig_dom_a}, the
following: on $u_U: u = 1$, on $u_D: u = 0$ and on $u_R: u^4 - u_0^4 +
\frac{\partial u}{\partial n} = 0$.  

\begin{figure}[h!]
  \centering
  \subfigure[]{\includegraphics[width=6.5cm,height=5.5cm]{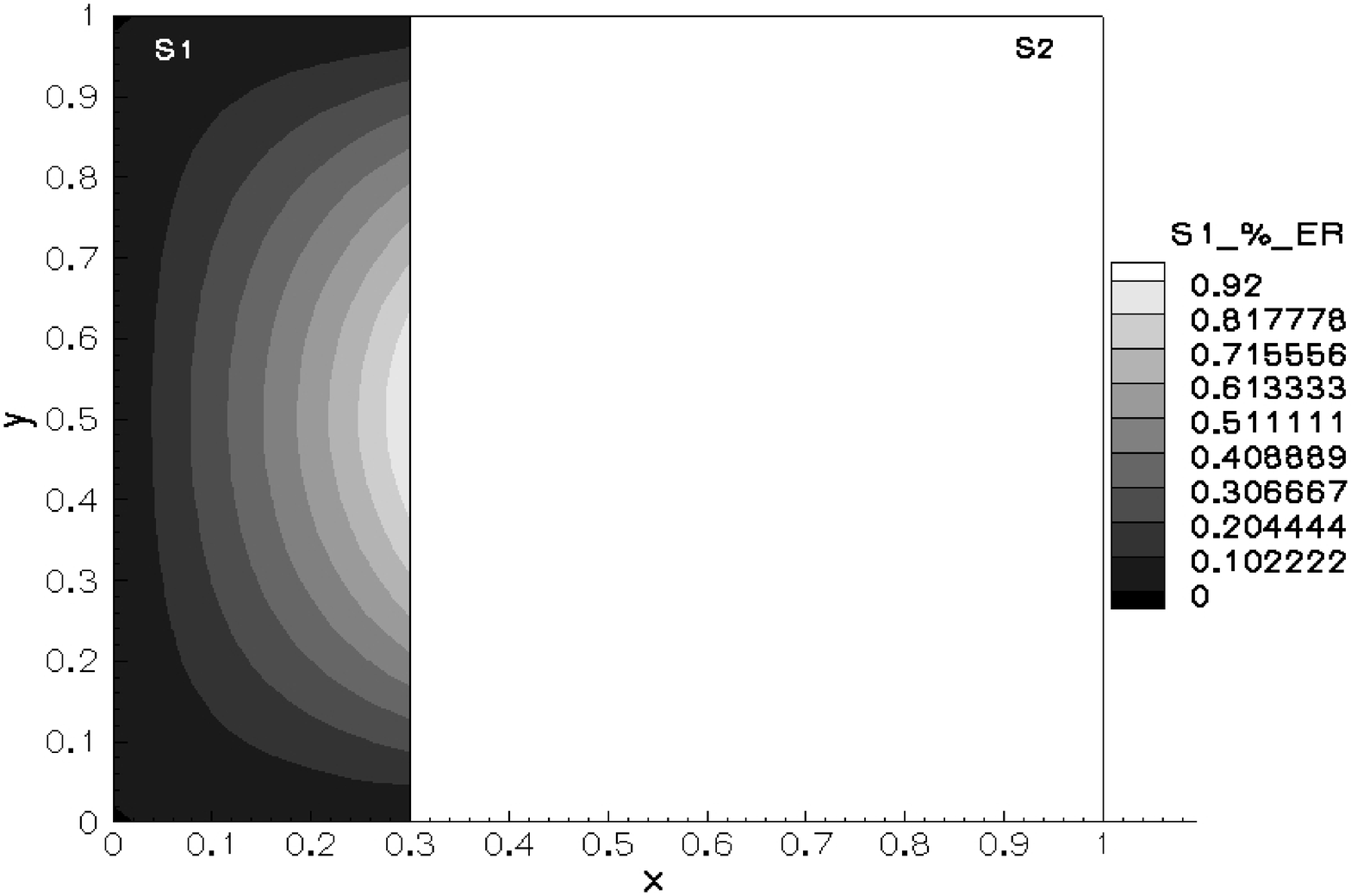}}
  \hspace{0.25cm}
  \subfigure[]{\includegraphics[width=6.5cm,height=5.5cm]{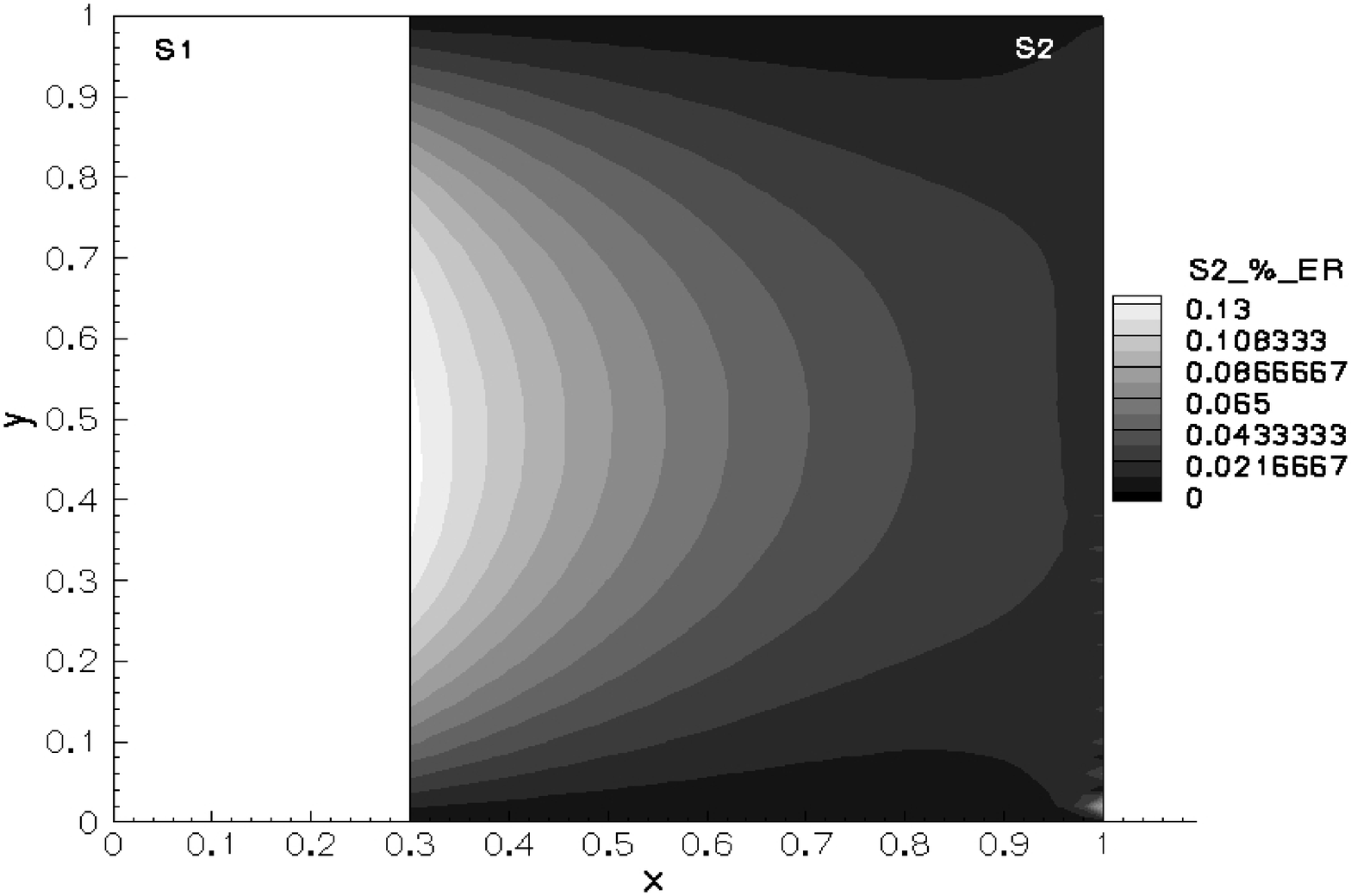}}
    \caption{\em Distribution of the relative error between the solution obtained
    by the present method and the solution obtained by a second order FD
    method on the whole domain. $\Omega_1$ (a), $\Omega_2$ (b).} 
    \label{fschur}
\end{figure}

The domain is split at $d = 1/3$. Then, the POD basis functions are 
generated on $\Omega_2$ using the previously computed database. In
order to check the accuracy of the method, a boundary condition which
was not included in the database used to build the POD modes  is
imposed on the left boundary: $u_L= y^2$, and a second order FD method
is used to solve the problem in $\Omega_1$. 
We use 6 POD modes to recover $\hat{u}_2$ inside
$\Omega_2$. Figure~\ref{fschur} presents the result of the test by
means of the distribution of the relative error between the FD
solution on the entire domain and the approximate Schur complement
approach. 

\subsubsection{Schwarz method}
\label{schw}

\begin{figure}[h!]
  \centering 
  \includegraphics[width=11.cm,height=6.cm]{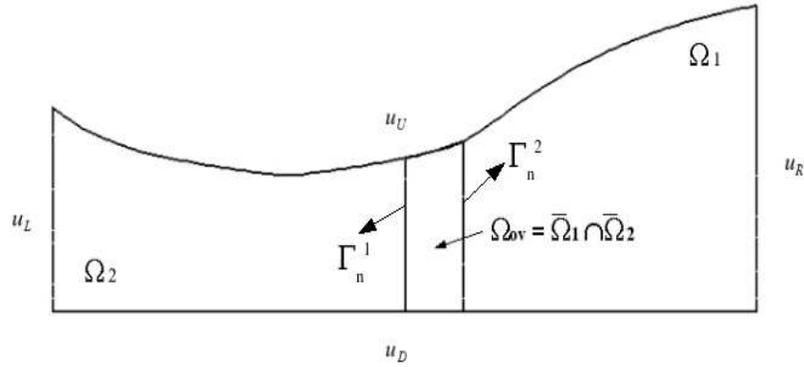}%
  \caption{{\em Problem set-up: Schwarz}}
  \label{fig_dom_b}
\end{figure}

In the following, a convergent-divergent  domain $\Omega$ is considered. 
As before, $\Omega$ is divided in two subdomains, $\Omega_1$
and $\Omega_2$, in a way that there exist an overlap region
$\Omega_{ov}=\overline{\Omega}_1 \cap \overline{\Omega}_2$ shared by
both subdomains (see fig.~\ref{fig_dom_b}). The Laplace
equation is solved with the boundary conditions detailed below to generate a
database of $k$ solutions with $1 \leq k \leq 60$, this time varying
the geometry of $\Omega_1$. The solution is obtained by means of the
finite element method as implemented in  ~\cite{ffem}, using P1 elements on a triangular
non-structured mesh and a fixed point iteration. The
boundary conditions imposed are: on $u_L: u = \frac{1}{3} \cdot y$, on
$u_U: u = 1$, on $u_D: u = 0$ and on $u_R: u^4 - u_0^4 +
\frac{\partial u}{\partial n} = 0$.   

For a geometry of the divergent part which is not included in the
database, the solution is determined following a similar approach to
that described for the Schur complement but employing this
time the classical Schwarz method (see, for example, \cite{Q99}):  
{\sf
\begin{enumerate}
\item
  solve the problem in $\Omega_1$ by any discretization method
  imposing Dirichlet b.c. on $\Gamma_n^1$;  
\item
  on $\Gamma_n^2$ project the trace of the above solution
  in the subspace spanned by the trace of the POD modes $\phi_i$;  
\item
  recover $\hat{u}_2$ as the prolongation of the trace of $\hat{u}_2$
  on $\Gamma_n^2$  inside $\Omega_2$ by using the POD modes;   
\item
  set $u_1=\hat{u}_2$ on $\Gamma_n^1$;
\item
  goto (1), $n = n + 1$, until convergence is attained.
\end{enumerate}}

\begin{figure}[h!]
  \centering
  \subfigure[]{\includegraphics[width=7.cm,height=7.cm]{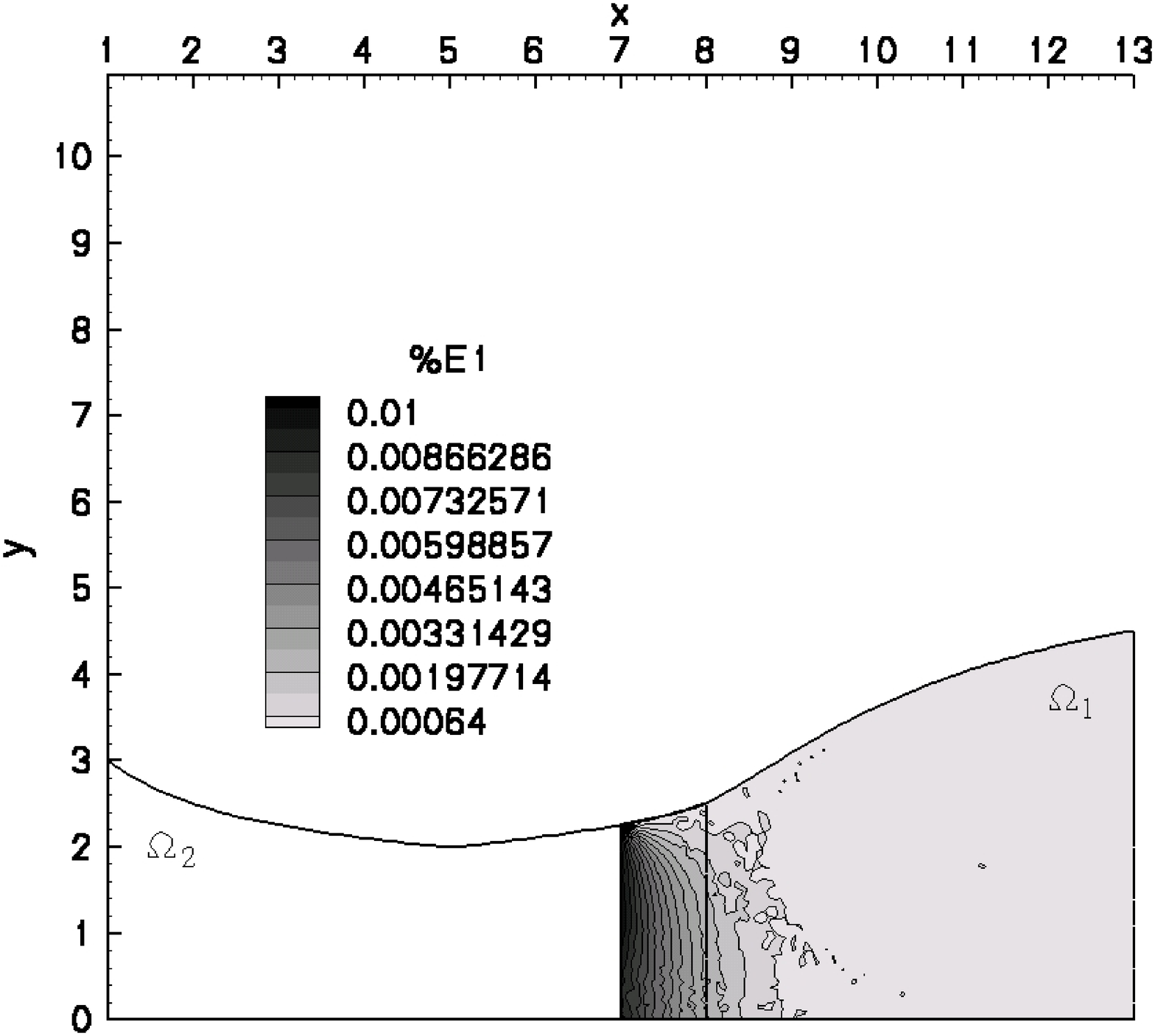}}%
  \subfigure[]{\includegraphics[width=7.cm,height=7.cm]{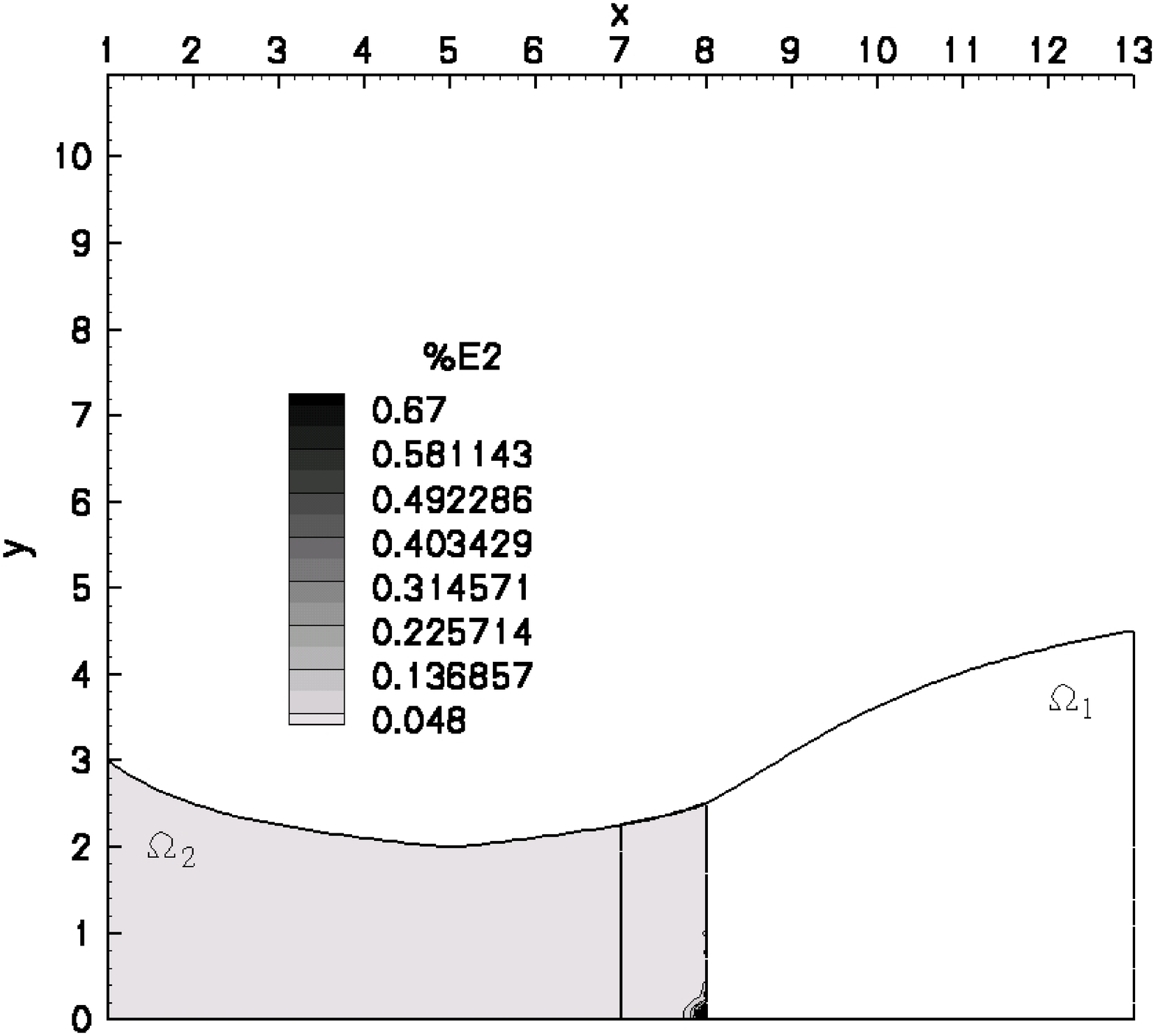}}%
  \caption{\em Percentage error distribution on (a) $\Omega_1$ and (b)
    $\Omega_2$. The reference solution is obtained by FEM P1 elements on a
    triangular mesh.}  
  \label{fschw}
\end{figure}

Four POD modes are used to recover $\hat{u}_2$ inside
$\Omega_2$. Figure~\ref{fschw} shows the results obtained for this
case, again in terms of the relative error between a numerical
solution (FEM P1) on the entire domain and the approximate Schwarz
method.       

\subsubsection{Newton method}
\label{nwt}

In this case the two-dimensional compressible Euler equations are solved
\begin{equation}
  \begin{array}{rcl} 
    0 & = & -  \vect{v} \cdot \dgrad S \\%
    0  & = & - \vect{v} \cdot \dgrad a - \frac{\gamma -1}{2}a \ddiv
    \vect{v} \\%
    0 & = & -\dgrad \vect{v}\cdot \vect{v} -
    a\left(\frac{2}{\gamma-1}\dgrad a - \frac{1}{\gamma(\gamma-1)} a
      \dgrad S \right),
  \end{array}
  \label{euler}
\end{equation}
with $\gamma = 1.4$ for air and being $S$ the entropy, $a$ the speed
of sound and $\vect{v} = [u,v]$ the velocity, in a
convergent-divergent nozzle on a structured mesh. We consider shockless flows. 

The $\lambda$-scheme
\cite{zaza} is used to solve the equations. Total temperature, total
pressure and the flow angle are imposed at the inlet; static pressure
at the exit and impermeability at the walls. The complete system is
solved by Newton iterations. The resulting linear problems are solved
by preconditioned GMRES iterations~\cite{spkit}.  

Using this code, a database of 90 snapshots is computed. 
The geometry of the divergent part of the nozzle is changed,
$\Omega_1$ in fig. \ref{fig_dom_b}. Furthermore, 
the static pressure at the exist is also varied 
taking uniformly spaced values in the interval $p = [0.94, 0.99]$ with
step $0.01$. This corresponds to 
15 snapshots for each pressure step. 
Since the flow is shockless, total temperature as well as total pressure are 
constant across the nozzle. Therefore, in order to solve in $\Omega_1$, the only 
unknown boundary condition is the flow angle on $\Gamma_n^1$. Hence, 
a low-order representation of  the flow angle $v/u$ for the convergent
part, i.e., $\Omega_2$,  is 
constructed retaining 10 POD modes.

We consider a case for which the geometry of the divergent part of the
nozzle and the static pressure imposed at the exit do not belong to
the set 
used to build POD database. The two sub-problems are coupled by a least
squares approximation on the overlapping domain. This is just a variant of the
Schwarz method presented in the previous section. The only difference
is that the two coupled problems are solved all at once by a
Newton method.      

The relative errors restricted to $\Omega_1$  are reported in table
\ref{tab1} for some of the flow variables. 
These errors are computed with respect to the numerical simulation on the entire
domain, and are quantified in terms of relative error in $L^2$ norm,
i.e., the $L^2$ norm of the difference between the solution obtained
by the present method and the reference solution divided by the $L^2$
norm of the reference solution. 

\begin{table}[h!]
\centering
\begin{tabular}{|c|c|c|c|c|c|}
\hline
Variable & $e(u)\%$ & $e(v)\%$ & $e(Ma)\%$ & $e(v/u)\%$\\%
\hline
Error & 1.1 & 2.25 & 1.14 & 0.75 \\%
\hline
\end{tabular}
\caption{Relative percentage errors (in $L^2$ norm) for the flow
  variables: $u$ (horizontal velocity), $v$ (vertical velocity), $Ma$
  (Mach number) and $v/u$ (flow angle).}  
\label{tab1}
\end{table}

\begin{figure}[h!]
  \centering
  \subfigure[]{\includegraphics[width=7.cm,height=6.5cm]{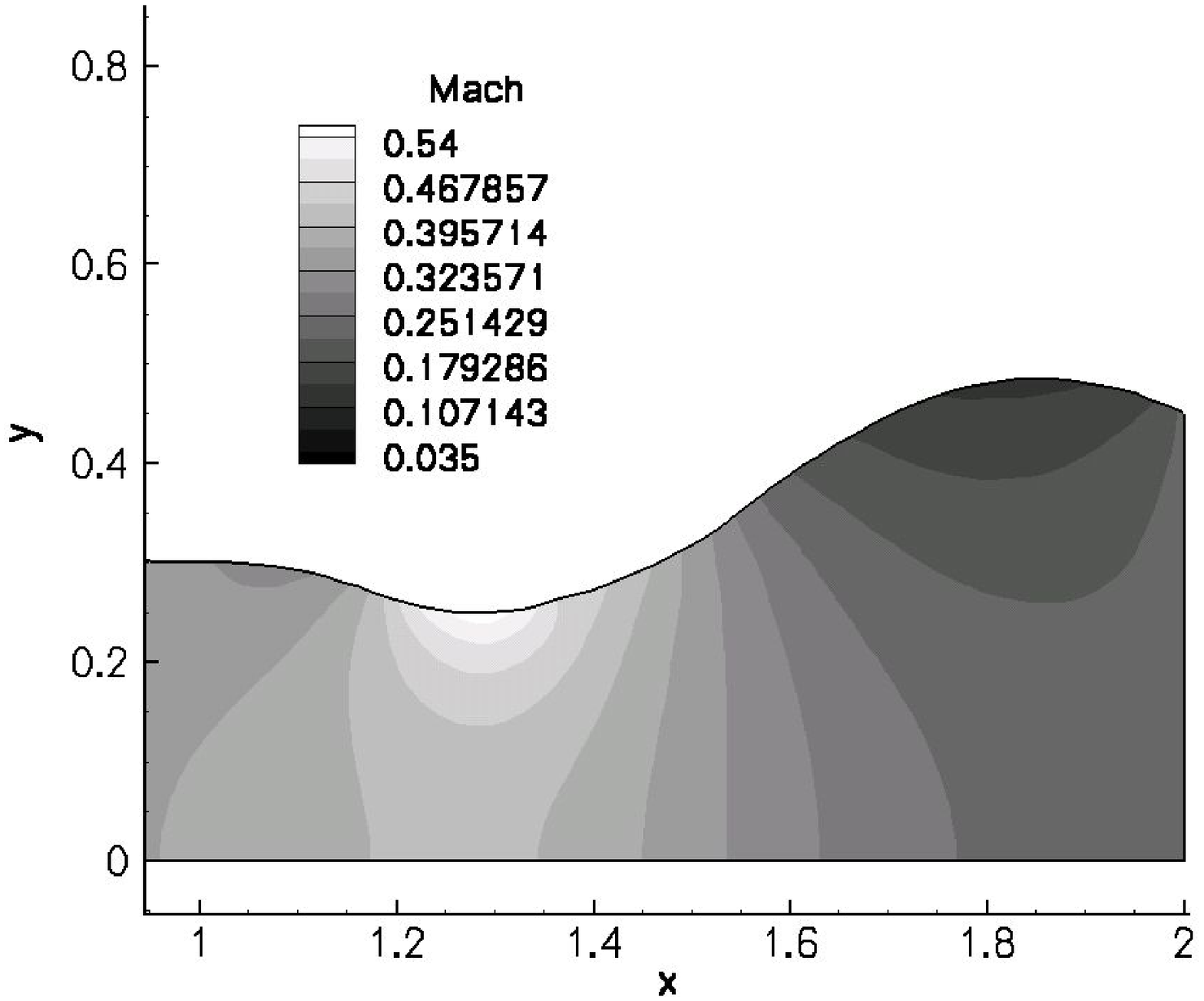}}%
  \subfigure[]{\includegraphics[width=7.cm,height=6.5cm]{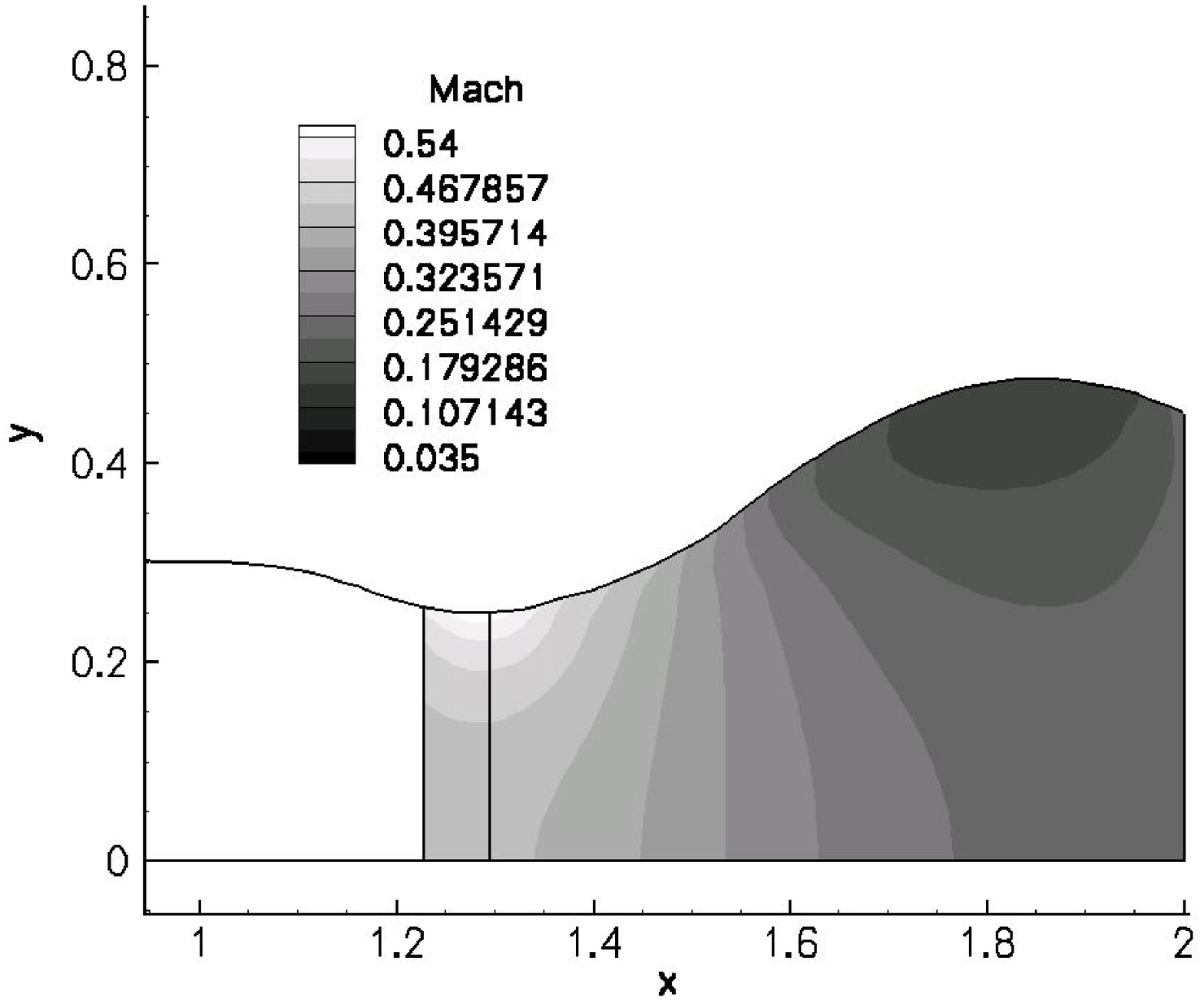}}\\%
  \subfigure[]{\includegraphics[width=7.cm,height=6.5cm]{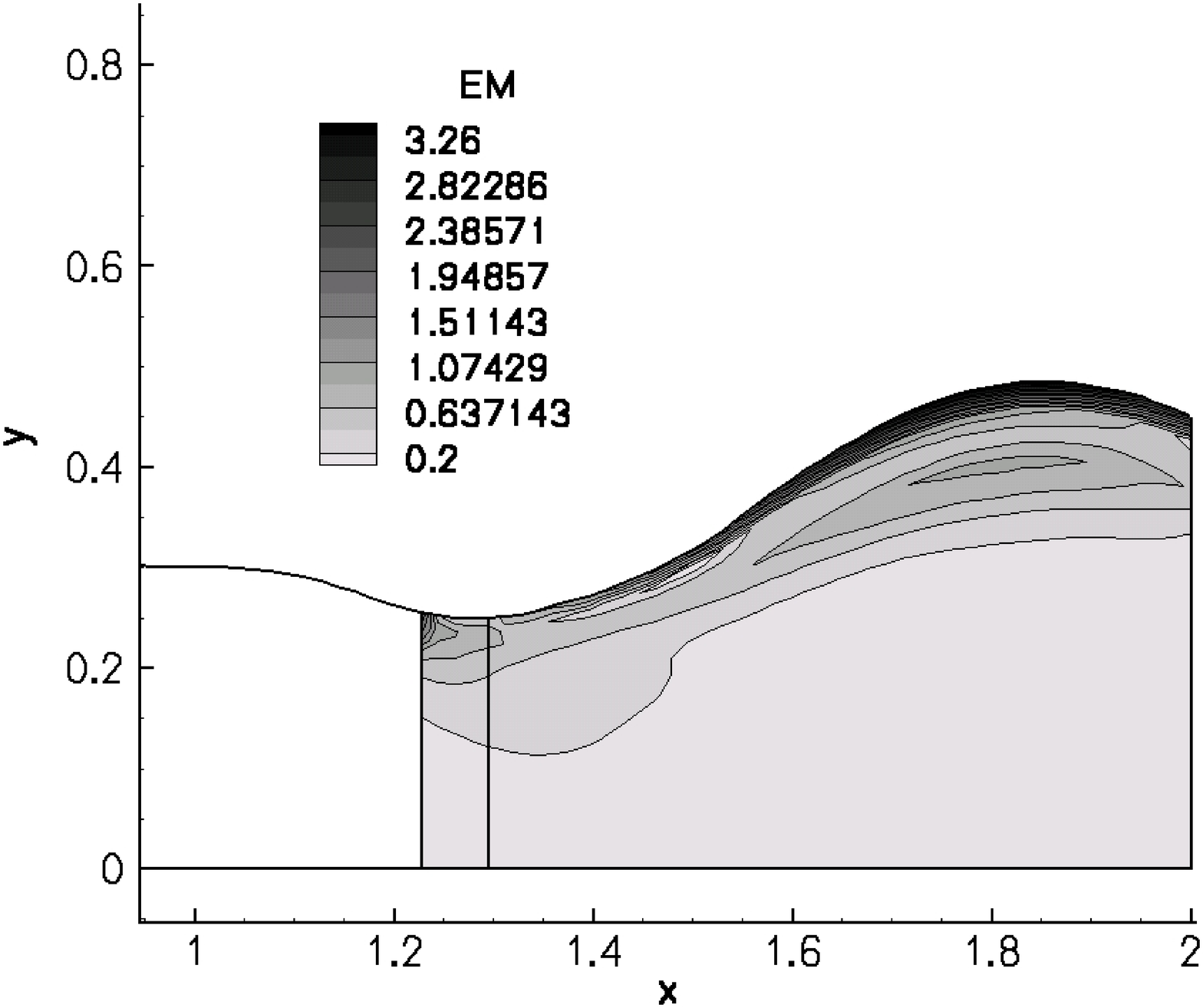}}%
  \caption{\em Distribution of $Mach$ number: (a) Numerical solution
    ($\lambda$-scheme). (b) Present method. (c) Relative percentage
    error distribution on $\Omega_1$.}     
  \label{feur_mach_1}
\end{figure}

\begin{figure}[h!]
   \centering
   \subfigure[]{\includegraphics[width=7.cm,height=6.5cm]{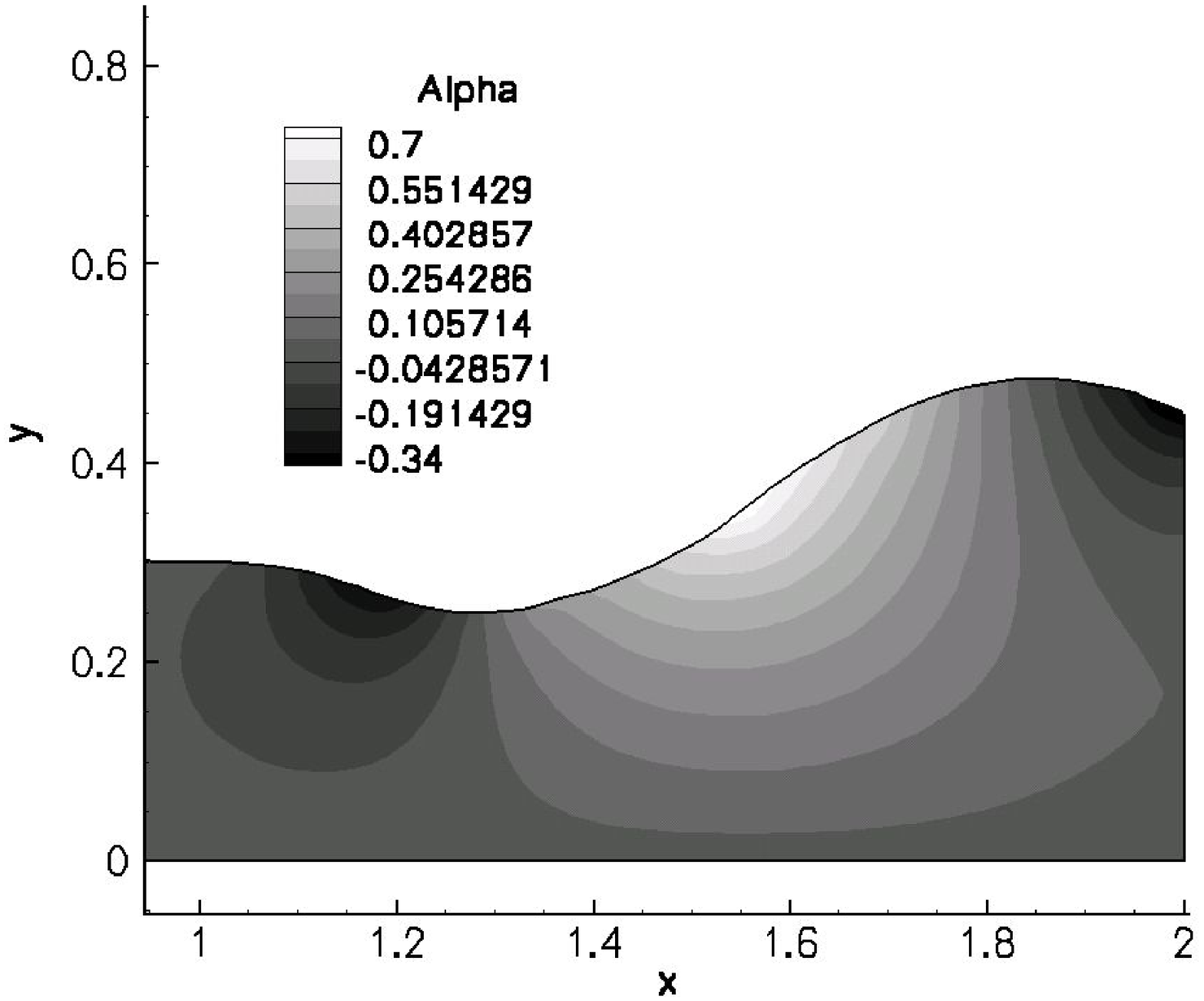}}%
   \subfigure[]{\includegraphics[width=7.cm,height=6.5cm]{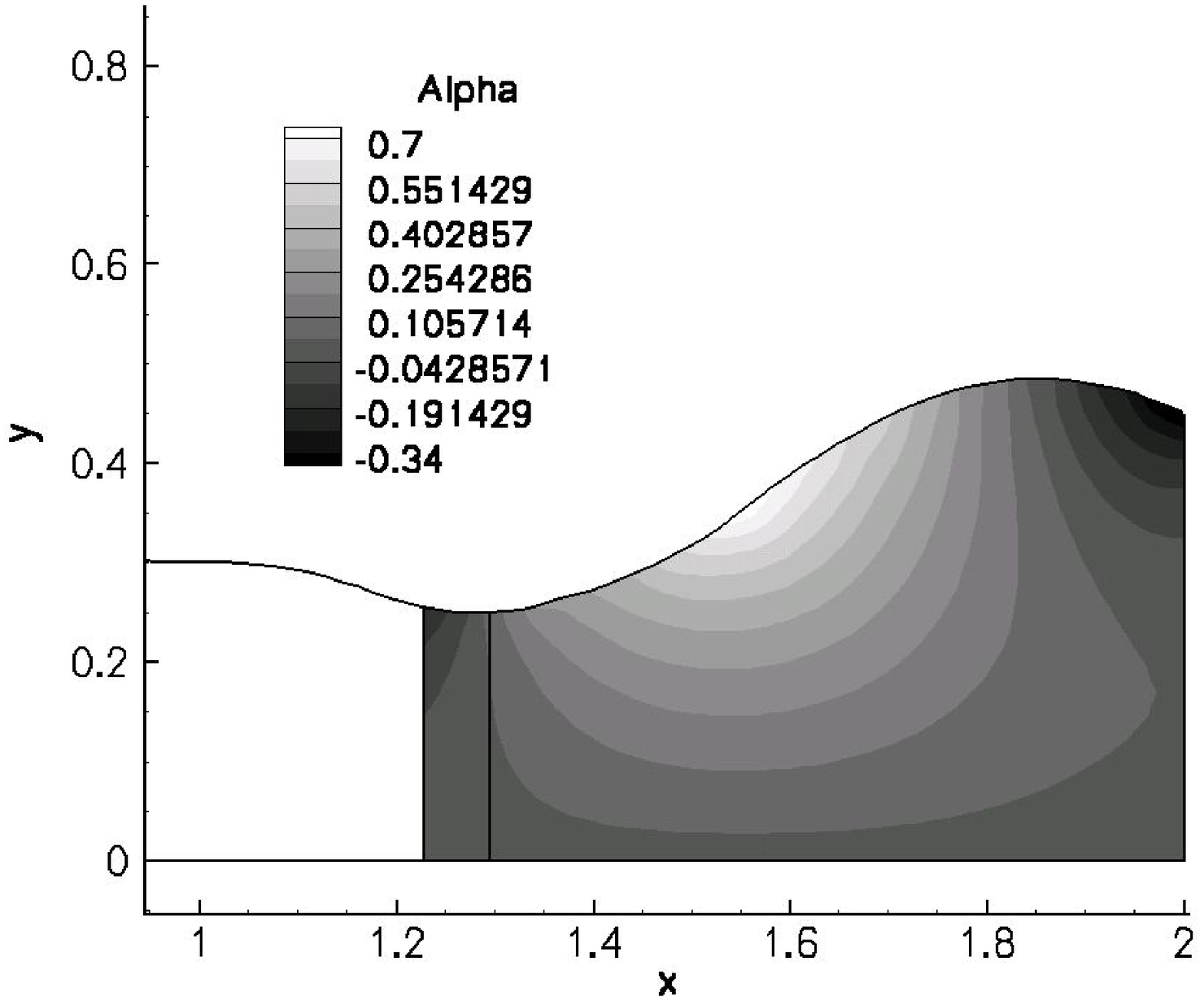}}\\%
   \subfigure[]{\includegraphics[width=7.cm,height=6.5cm]{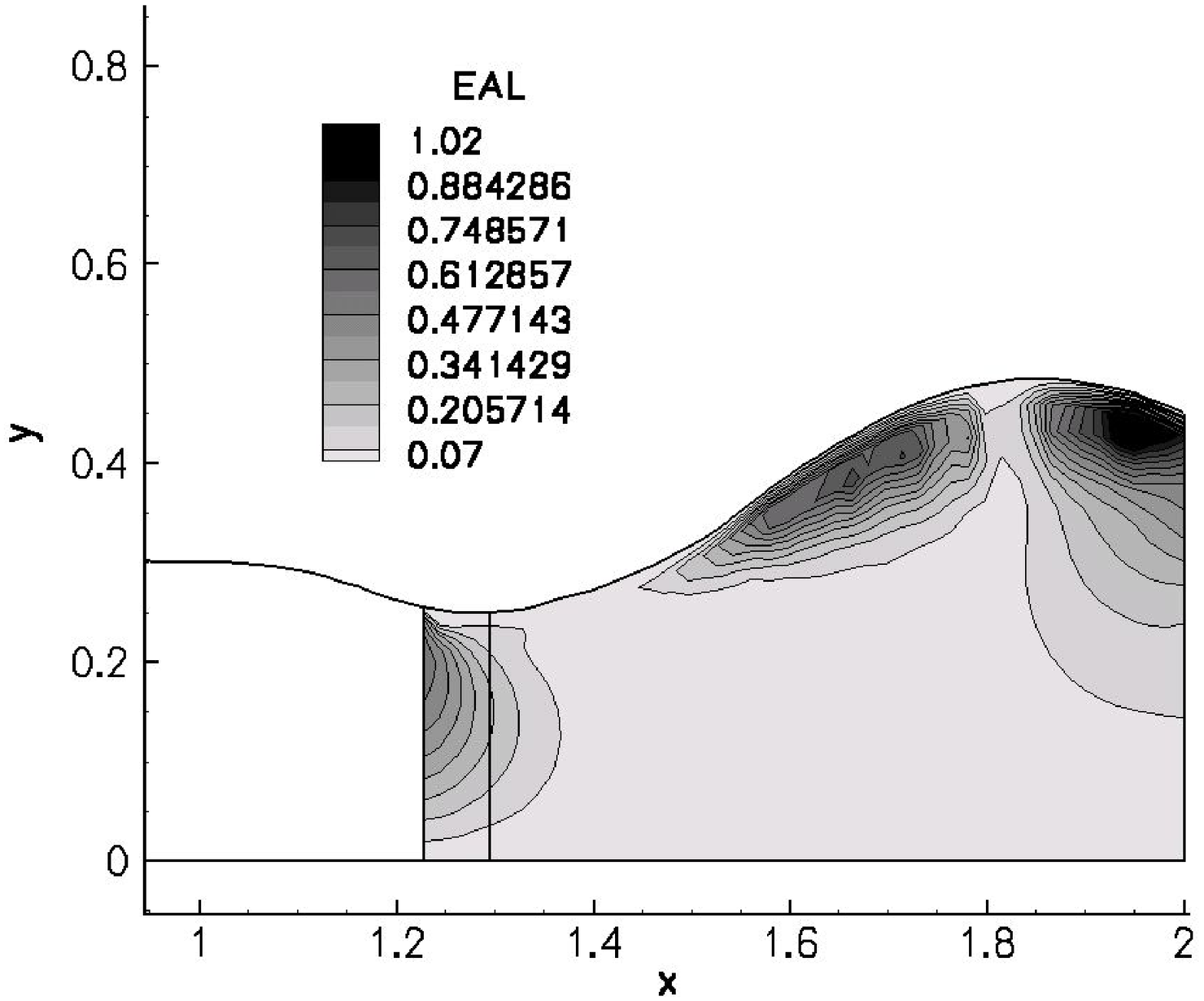}}%
   \caption{\em Distribution of $v/u$. (a) Numerical solution
     ($\lambda$-scheme) on the entire domain. (b) Present method. (c)
     Relative percentage error distribution on $\Omega_1$.}   
   \label{feur_alph_1}
 \end{figure}

Figure \ref{feur_mach_1}(a) presents the distribution of the Mach
number obtained by the full numerical simulation on the entire
domain. The same quantity  obtained when applying the
proposed method is shown in fig. \ref{feur_mach_1}(b). Finally, 
fig. \ref{feur_mach_1}(c) shows the distribution of the relative
percentage error. In figure~\ref{feur_alph_1} the flow angle $v/u$ is considered.

\section{Solution by minimization of the residual norm in the space
  spanned by the POD modes} 
\label{sec:3}

An alternative way to couple the low-order model to a detailed
simulation is to look for an approximate solution in the reduced order
function space that takes into account the governing equations. 
Hence, the main difference with respect to the approach described in
section \ref{sec:2} is that the approximate solution is
found by minimizing the residual norm of a given discretization scheme
on the whole domain $\Omega_2$ rather than by projecting the trace of
the solution in the space spanned by POD modes.  

Again the total pressure and the total temperature are constant across the nozzle and 
therefore we can write any other variable as a function of the local Mach number and 
the ratio $v/u$. In particular let $U$ be the array of the couples $(Ma,v/u)$ for all 
the grid points belonging to $\Omega_2$.   
We start by representing this vector in the original
$Q$-dimensional discrete space by a linear combination of basis functions, $U
= \sum_{i=1}^M \alpha_i  \Phi_i$, with  $M\ll Q$. The arrays $\Phi_i$ 
have the same structure of $ U$ and they have been obtained by POD. The idea is to 
satisfy the compressible Euler equations (\ref{euler}) in a least squares
sense over $\Omega_2$. In other words, let $E( \alpha)$ be the discrete residual 
of the governing equations as a function of $ \alpha=(\alpha_1,\dots,\alpha_M)$
and let 
\begin{equation}
  I( \alpha) = \frac{1}{2} E^T( \alpha)  E(
  \alpha) 
\end{equation}
be the residual norm. The solution in the POD function space  
$ \alpha^*$ is found by setting 
\begin{equation}
   \alpha^* = \textup{arg min}_{ \alpha} I( \alpha)
\end{equation}
This is equivalent to a system of non-linear equations 
\begin{equation}
  \frac{\partial I}{\partial \alpha_i} = \frac{\partial  E^T(
    \alpha^* )}{\partial \alpha_i}  E( \alpha^*) = 
  J^T( \alpha^*)  E( \alpha^*) = 0 \quad \forall i \in [1,\dots,M]
  \label{eq:1}
\end{equation}
where $ J^T$ is the system's Jacobian.

Equation (\ref{eq:1}) is solved by a quasi-Newton method where we take 
\begin{equation}
  \begin{array}{rcl} 
     E( \alpha^*) & = &  E( \alpha^0)
    +  J( \alpha^0) \Delta  \alpha + O(\Delta 
    \alpha^2)
  \end{array}
\end{equation}
and after substituting in \eqref{eq:1} we obtain
\begin{equation}
   J^T  J \Delta \alpha_i = - J^T  E
  \label{eq:2}
\end{equation}
Therefore, we are left with the solution of equation (\ref{eq:2}) at
each step of the quasi-Newton algorithm employed to solve (\ref{eq:1}).

Using the same database generated in Section \ref{nwt}, two low-order
basis restricted to the convergent part of the
nozzle are constructed. One for the flow angle $v/u$ and the
other for the Mach number. Only 5 POD modes are retained for both the
flow angle and the Mach number.

\begin{figure}[h!]
  \centering
  \subfigure[]{\includegraphics[width=7.cm,height=6.5cm]{./FIGURES/alpha_sntot_111_0965.eps}}%
  \subfigure[]{\includegraphics[width=7.cm,height=6.5cm]{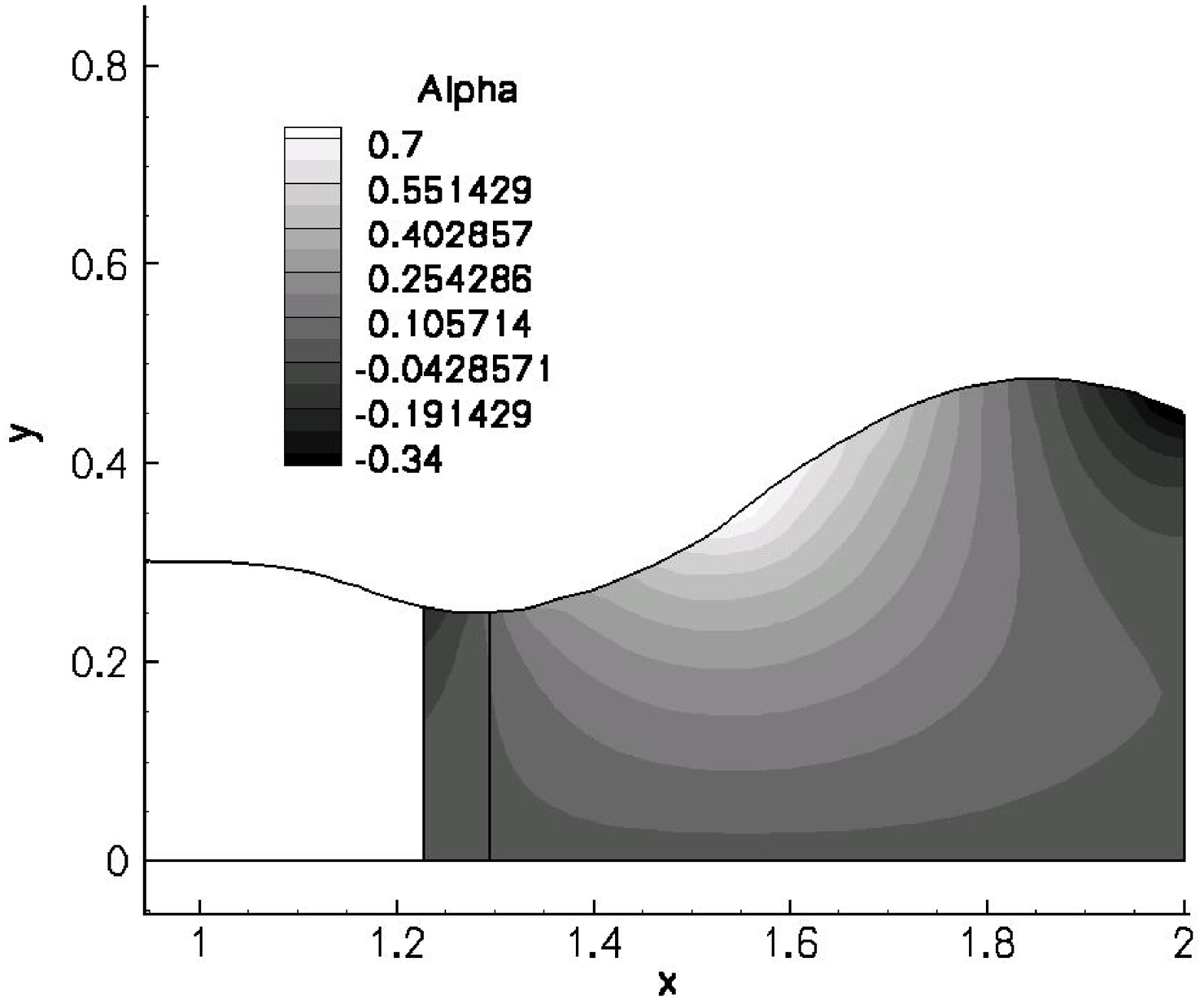}}\\%
  \subfigure[]{\includegraphics[width=7.cm,height=6.5cm]{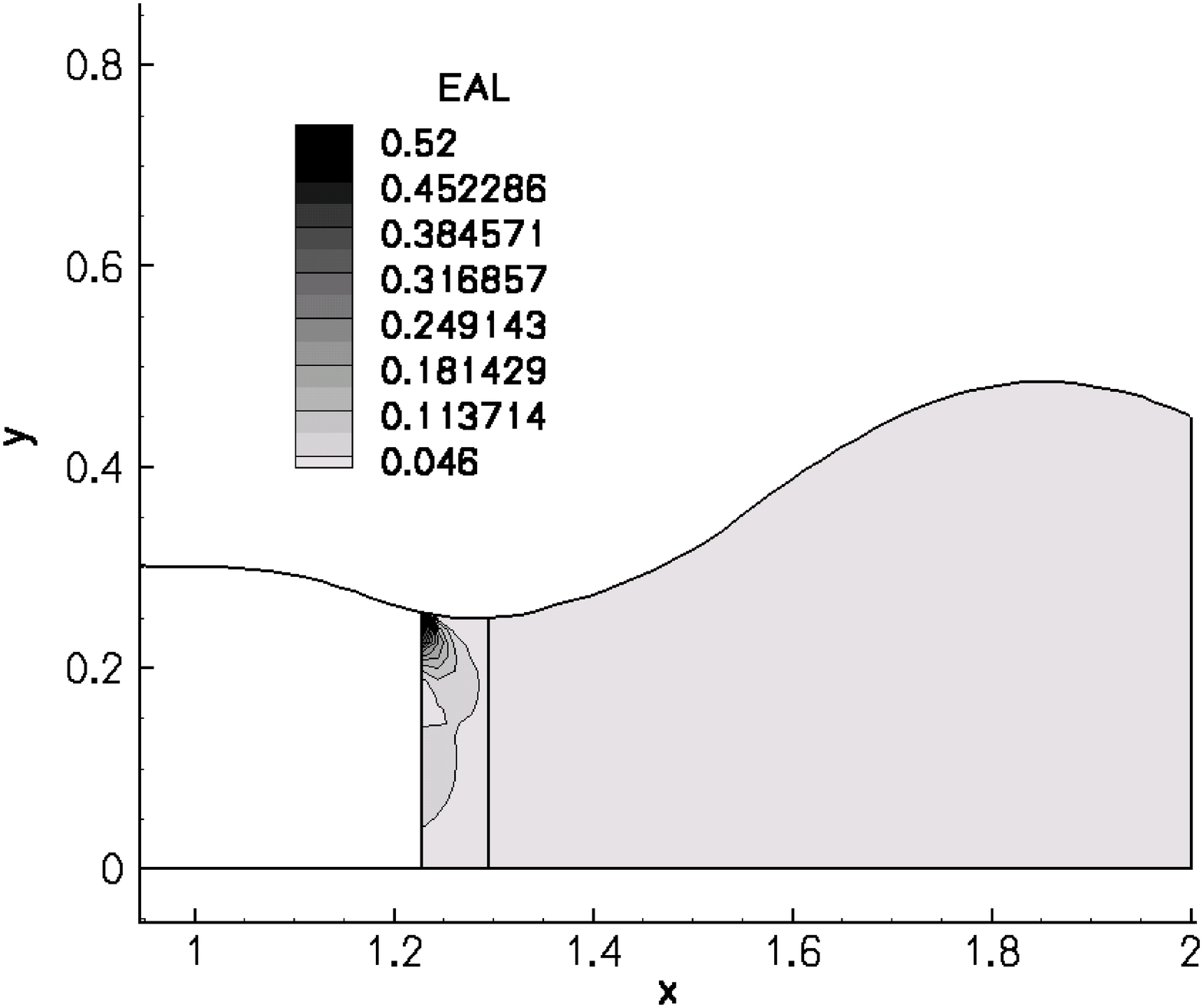}}%
  \caption{\em Distribution of $v/u$. (a) Numerical solution
    ($\lambda$-scheme) on the entire domain. (b) Present method. (c)
    Relative percentage error distribution on $\Omega_1$.}  
  \label{feur_alph_2}
\end{figure}

This method is tested for the same case as for the Newton
method presented in Section \ref{nwt}. The only difference is that the
solution in $\Omega_2$ is found by
minimization of the residuals norm in the space spanned by the POD
modes. Otherwise,  the full
numerical simulation is employed on $\Omega_1$, and 
a classical  Schwarz overlapping method is used to iterate the
solution to convergence.

\begin{figure}[h!]
  \centering
  \subfigure[]{\includegraphics[width=7.cm,height=6.5cm]{FIGURES/mach_sntot_111_0965.eps}}%
  \subfigure[]{\includegraphics[width=7.cm,height=6.5cm]{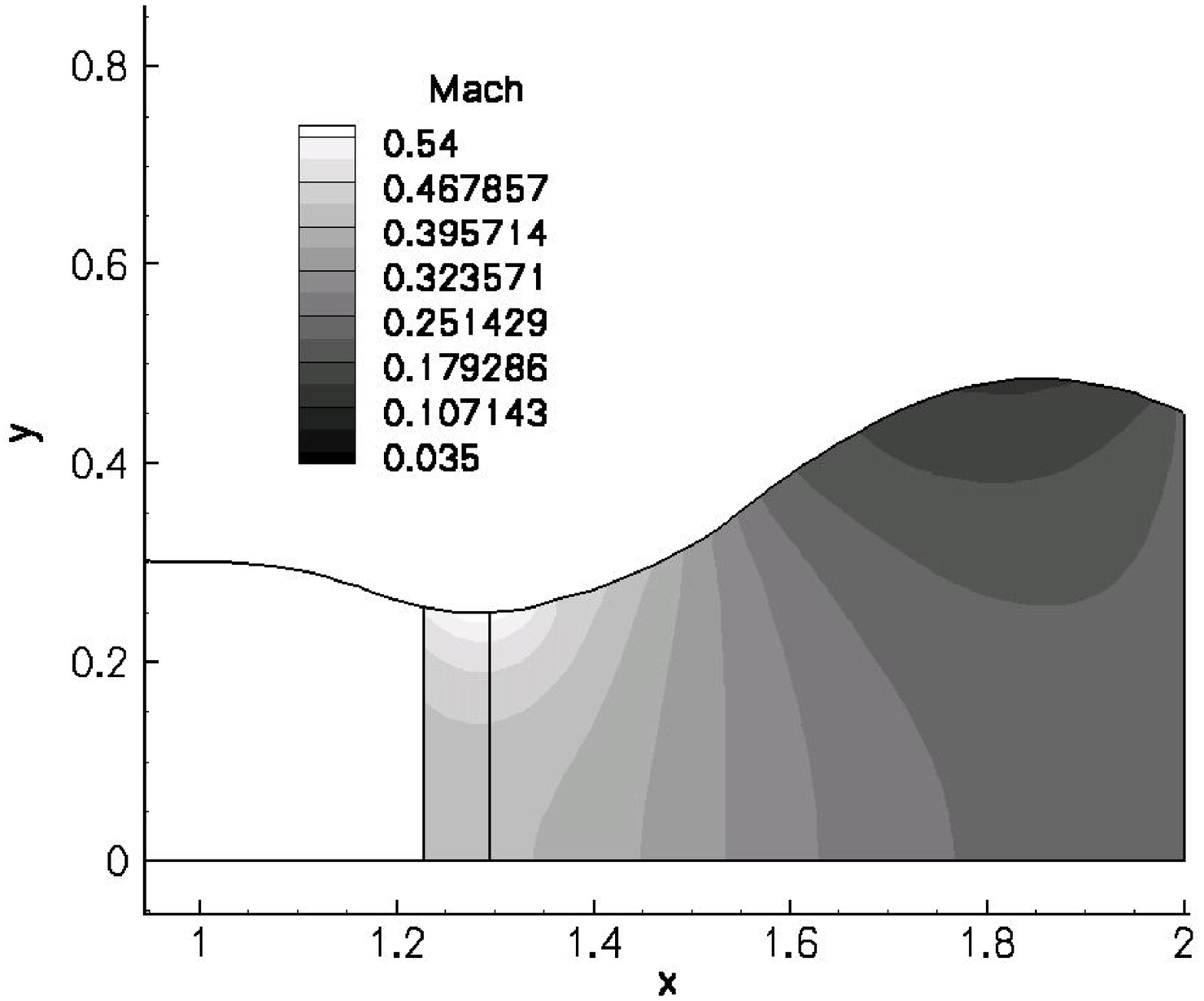}}\\%
  \subfigure[]{\includegraphics[width=7.cm,height=6.5cm]{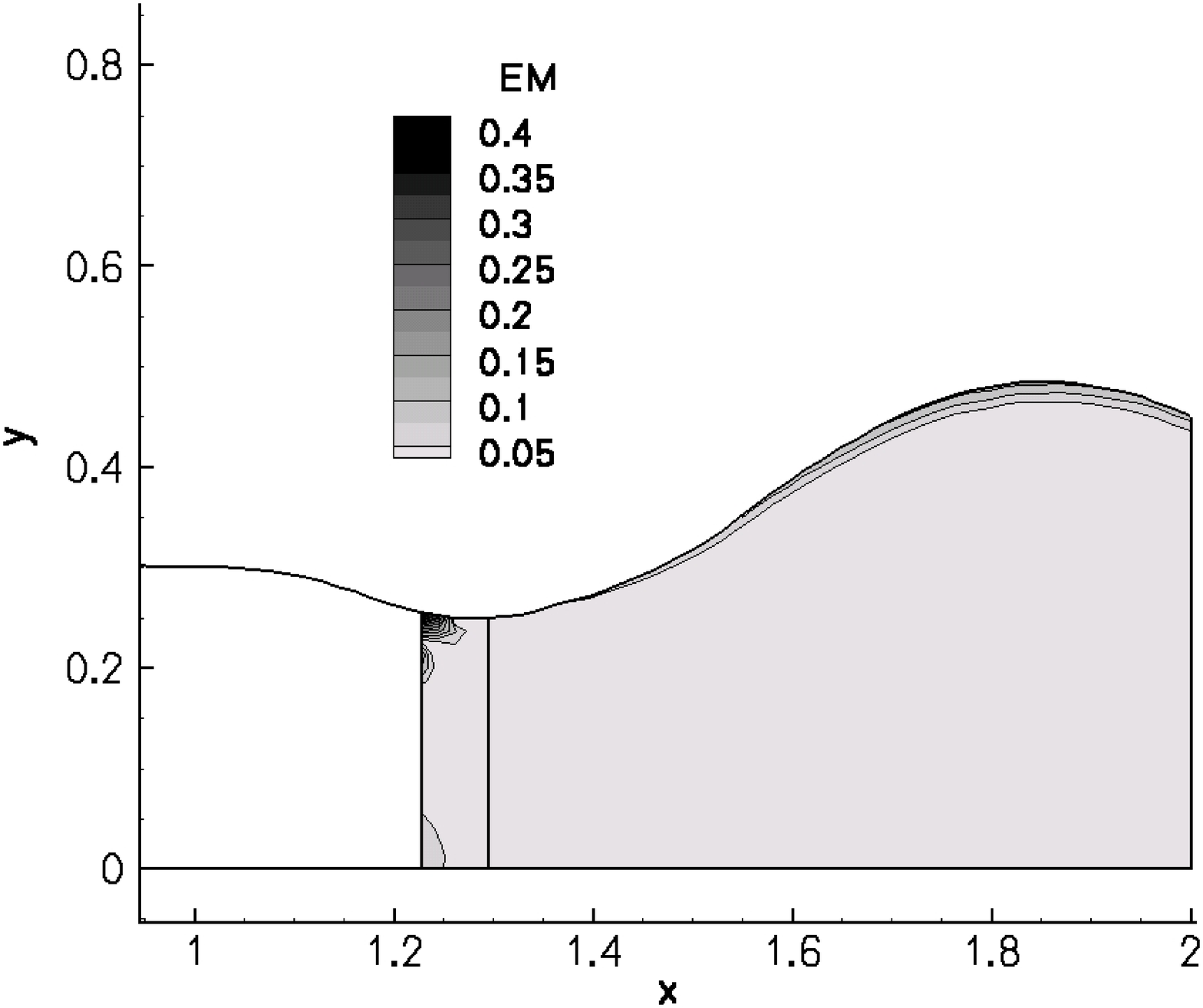}}%
  \caption{\em Distribution of $Macxh$ number on $\Omega_1$. (a)
    Numerical solution ($\lambda$-scheme). (b) Present method. (c) 
    Relative percentage error distribution on $\Omega_1$.}   
  \label{feur_mach_2}
\end{figure}

The reference Mach number on the entire domain, the Mach
number obtained by the present method on $\Omega_1$ as well as the
distribution of the relative percentage error on $\Omega_1$ are shown
in figures \ref{feur_mach_2}(a), \ref{feur_mach_2}(b) and
\ref{feur_mach_2}(c), respectively.  

In fig. \ref{feur_alph_2}, the results for the flow angle $v/u$ are
shown while table \ref{tab2} presents the relative errors obtained in  $\Omega_1$.    
\begin{table}[h!]
\centering
\begin{tabular}{|c|c|c|c|c|}
\hline
Variable & $e(u)\%$ & $e(v)\%$ & $e(Ma)\%$ & $e(v/u)\%$\\%
\hline
Error & 0.05 & 0.26 & 0.05 & 0.13 \\%
\hline
\end{tabular}
\caption{Relative percentage errors (in $L^2$ norm) for the flow
  variables: $u$ (horizontal velocity), $v$ (vertical velocity), $Ma$
  (Mach number) and $v/u$ (flow angle).}  
\label{tab2}
\end{table}

These results show a much better accuracy as compared to those of section
\ref{nwt} as one can conclude by comparing tables \ref{tab1} and
\ref{tab2}. 

Finally, in order to asses sensitivity of the results 
with respect to the minimization problem, we show in table \ref{tab-nuova1}
the error in $\Omega_2$ when we use for $\alpha$  the initial
guess. In table \ref{tab-nuova2}
we present the error at the end of the Schwarz iteration. It is seen 
\begin{table}[h!]
\centering
\begin{tabular}{|c|c|c|c|c|}
\hline
Variable & $e(u)\%$ & $e(v)\%$ & $e(Ma)\%$ & $e(v/u)\%$\\%
\hline  
Error & 35.52 & 45.65 & 39.28 & 6.25 \\%
\hline
\end{tabular}
\caption{Relative percentage errors (in $L^2$ norm) for the flow
  variables: $u$ (horizontal velocity), $v$ (vertical velocity), $Ma$
  (Mach number) and $v/u$ (flow angle).}  
\label{tab-nuova1}
\end{table} 
that the fact of approximating the solution of the Euler equations
with appropriate boundary conditions,
even in a crude way, in $\Omega_2$, significantly improves the
solution with respect to the initial guess.
\begin{table}[h!]
\centering
\begin{tabular}{|c|c|c|c|c|}
\hline
Variable & $e(u)\%$ & $e(v)\%$ & $e(Ma)\%$ & $e(v/u)\%$\\%
\hline
Error & 2.01 & 3.94 & 2.08 & 2.90 \\%
\hline
\end{tabular}
\caption{Relative percentage errors (in $L^2$ norm) for the flow
  variables: $u$ (horizontal velocity), $v$ (vertical velocity), $Ma$
  (Mach number) and $v/u$ (flow angle).}  
\label{tab-nuova2}
\end{table}
   
\section{Discussion}
\label{sec:4}

The fact of reducing the extent of the computational domain does not
guarantee that one can get a solution faster as compared to solving 
the problem on the entire domain. In particular, let us consider the
method described in section \ref{nwt}. It would in principle be the
less demanding in terms of computational time since it just amounts to a
non-local boundary condition in the frame of a Newton
method. However, this is
not necessarily the case since the convergence of the implicit iteration is spoiled
by such non-local boundary condition and the number of Newton steps 
to attain convergence goes from 7 on the whole domain, to 11
when we only solve on $\Omega_1$.  The slower convergence rate may lead
to comparable costs in terms of CPU time, as a function of the
fraction of the domain that is actually resolved. On the other hand,
the fact of reducing the number of grid points is reflected almost
proportionally on the memory requirements. However, the
Jacobian matrix will have a non-sparse block corresponding to the
non-local boundary condition induced by the approximation of the
Steklov-Poincar\'e operator. 

Concerning the method described in
section  \ref{sec:3} the cost of each step of the
proposed minimization algorithm can be split up as follows. 
i) The computation of $ J$. The Jacobian is evaluated by
  one-sided finite differences. Consequently its cost in terms of CPU
  time and memory requirements is proportional to the dimension of
  the low-order space $M$ and the cost of computing the residual
  vector, i.e.,  $\ct( J) = M \times \ct( E)$. Since $M$ is
  $O(10)$, the Jacobian matrix can be formed after few residual
  evaluations that are cheap in terms of CPU time. Memory
  requirements could become prohibitive for very big problems only on  
  serial architectures.   
ii)  The computation of the symmetric matrix $ A = J^T
   J$. The dimensions of $ A$ are  $M \times M$, with
  $M=O(10)$. 
  The required floating point operations are $(M \times
  M)/2 \times N$ and they are less than the operations needed for the
  computation of $ E$.  iii)  
  The cost of finding the solution to the 
  linear system of size $M \times M$ can be neglected.  
Summarizing, it can be deduced that the cost is equivalent
to some residual evaluations. The number of iterations needed in
order to find the minimum are generally less than five, yielding an
overall cost negligible with respect to canonical CFD calculation on $\Omega_2$.

We remark two major
limits of these approaches. The first is of course that the results
depend to a large extent on the database used for the POD modes. If
the configuration under consideration lies in a region of the
parameter space far from that explored when building the database,
then the approximation error can be large. It is true, however, that the
low-order model should be used where the solution does not strongly depend
  on the boundary conditions or the geometry. Another
limitation is that an efficient way to improve the approximation
quality comparable for example to grid refinement is not available. In
principle, we would like to increase the approximation accuracy by
enriching the functional space in which the solution is sought, based on
some objective criteria. Unfortunately a general framework for such 
improvement is not presently available and it is the object of
present research.

In conclusion we presented some possible implementations of a method to reduce the
extent of the computational domain in the numerical solution of
partial differential equations. The idea of using models that take into
account different physical phenomena in different subdomains is 
old. Here we revisited this approach using a low-order model in the
framework of classical domain decomposition techniques. The
results in terms of the approximation error are promising for all of the
cases that we showed. A major challenge to be pursued is to find  a
viable a posteriori error estimation technique for
iteratively adapting the POD function space. 

\bibliography{rr_bti}
\bibliographystyle{plain}
\end{document}